\documentclass[epj]{svjour}

\usepackage{graphicx}
\usepackage{fancyhdr}
\usepackage{amsfonts,amsmath,fixmath}
\def\makeheadbox{{
 }}
\def\mytitle{My title} 
\def\myauthors{My name}  
\def\mytype{My type of session}
\def\mysession{My session}

\def\mytitle{Searches for leptoquark production at D0} 
\def\myauthors{Thomas Nunnemann}    
\def\mytype{Contributed Talk}    
\def\mysession{Alternatives}

\def\gsim{\,\lower.25ex\hbox{$\scriptstyle\sim$}\kern-1.30ex%
\raise 0.55ex\hbox{$\scriptstyle >$}\,}
\def\lsim{\,\lower.25ex\hbox{$\scriptstyle\sim$}\kern-1.30ex%
\raise 0.55ex\hbox{$\scriptstyle <$}\,}

\newcommand{\gev}{\ensuremath{\mathrm{GeV}}}

\newcommand{\tev}{\ensuremath{\mathrm{TeV}}}

\newcommand{\MET}{\mbox{\ensuremath{E \kern-0.6em\slash_{\rm T}}}}
\newcommand{\MHT}{\mbox{\ensuremath{H \kern-0.75em\slash_{\rm T}}}}

\newcommand{\ra}{\ensuremath{\rightarrow}}
\newcommand{\pb}{\ensuremath{\mathrm{pb}}}

\newcommand{\pbi}{\ensuremath{\mathrm{pb}^{-1}}}
\newcommand{\fbi}{\ensuremath{\mathrm{fb}^{-1}}}

\begin{document}
\title{Searches for Leptoquark Production at D0}
\author{Thomas Nunnemann on behalf of the D0 Collaboration
\thanks{\emph{Email:} Thomas.Nunnemann@lmu.de}
}                     
%
%
\institute{Department f\"{u}r Physik, Ludwig-Maximillians Universit\"{a}t M\"{u}nchen, Am Coulombwall 1, D-85748 Garching, Germany}
%
\date{Received: 30.09.2007}
\abstract{Recent searches for leptoquark production in $p\bar{p}$ collisions
at $\sqrt{s}=1.96\,\tev$ are presented using data samples with integrated 
luminosities up to $1\,\fbi$ recorded with the D0 detector.
} 
\maketitle

\section{Introduction}
Leptoquarks, 
hypothetical colored bosons which carry both lepton and quark
quantum numbers and thus allow lepton-quark transitions, are predicted
by numerous extensions of the standard model (SM)~\cite{lqpheno}. At hadron 
colliders,
leptoquarks are predominantly produced in pairs via the strong coupling.
Single leptoquarks can be produced via $t$-channel leptoquark exchange,
which depends on the unknown leptoquark-lepton-quark coupling $\lambda$.

As the pair-production of scalar leptoquarks is a pure QCD 
process\footnote{when
neglecting the contribution from $t$-channel lepton exchange which is 
$\propto \lambda^2$} which has been calculated to NLO \cite{Kramer:1997hh},
the cross-section depends only on the assumed leptoquark mass $M_{LQ}$.
In case of vector leptoquarks, the pair-production cross section is generally
much larger and additionally depends on unknown anomalous couplings.
Furthermore, the cross section has only been calculated at LO. Therefore,
all searches by the D0 experiment presented below assume scalar leptoquarks.

Leptoquarks could, in principle, decay into any combination of a quark and a lepton,
but leptoquarks with masses as low as $\mathcal{O}(100\,\gev)$ are only allowed
to couple
to one generation of quarks and leptons, since they otherwise would generate
lepton number violation and sizable flavor-changing neutral currents.
The branching fractions of the leptoquark decays into a charged lepton and
quark or neutrino and quark are determined by the respective
$LQ$-$\ell$-$q$ coupling. Thus, leptoquark pair-production can produce
three characteristic final states: $\ell^+\ell^- qq$, $\ell^\pm\nu qq$, and
$\nu\nu qq$.

\section{Pair-production of second generation scalar leptoquarks}
Second generation scalar leptoquarks ($LQ_2$) were already searched for 
in the channel $LQ_2\overline{LQ}_2\ra \mu q \mu q$ using an integrated
luminosity of $250\,\fbi$~\cite{Abazov:2006vc}.
In combination with previous D0 measurements
lower mass limits of $M_{LQ_2}> 251\,\gev$ for 
$\beta = \mathcal{B}(LQ_2\ra \mu q)=1$ and $M_{LQ_2}> 204\,\gev$
for $\beta=1/2$ were set. All limits reported in this note are at 95\%
confidence level.

The first D0 search performed in Run\,II in the channel 
$LQ_2\overline{LQ}_2\ra \mu q \nu q$,
which has maximal sensitivity for $\beta=1/2$, is based on the Run\,IIa data
set of $1\,\fbi$~\cite{d05370}.
The preselection required
exactly one muon with large transverse momentum, $p_T>20\,\gev$, 
reconstructed in a wide range of pseudorapidity, $|\eta|<2$,
large missing transverse energy, $\MET>30\,\gev$, and at least two jets with
transverse energies $E_T>25\,\gev$ and $|\eta|<2.5$. 
In addition, events with the \MET{} direction
opposite in azimuth to the muon were removed, as they were likely due to
badly reconstructed muons resulting in an overestimated \MET{}.
The background consisted of $W$ + jets, $Z$ + jets (with a non-identified
$\mu$) and $t\bar{t}$-production, which were estimated from simulation, 
and a small contribution of QCD multijet production, which was derived from
data. The $W$ + jets background was normalized to data at
preselection level within a region of transverse mass $M_T(\mu\nu)$
dominated by $W$ boson production.

The leptoquark signal was discriminated from the background using the
muon-neutrino transverse
mass $M_T(\mu\nu)$, the scalar transverse energy 
$S_T=E_T(\mu)+\MET + E_T(\mathrm{jet}_1) + E_T(\mathrm{jet}_2)$, 
the transverse mass
$M_T(\nu j_1)$ constructed from \MET{} and the momentum of the leading jet, and
the invariant
mass of the muon jet combination closest to the assumed leptoquark mass. 
These
selection requirements are motivated by the high leptoquark mass and
consequently high transverse momenta of its decay products.
For an assumed leptoquark mass $M_{LQ_2}=200\,\gev$, six data events were
selected with a background prediction of $6.4\pm 1.1$ events, of which
50\% were from $W$ + jets production and the remainder from $t\bar{t}$,
$Z/\gamma^*$ + jets, and multijet production.
The systematic error on the $W$ + jet background prediction was
found to be dominated by uncertainties in the jet energy scale and in the
modeling of the jet transverse momentum shapes. 

\begin{figure}
\begin{center}
\includegraphics[width=0.45\textwidth]{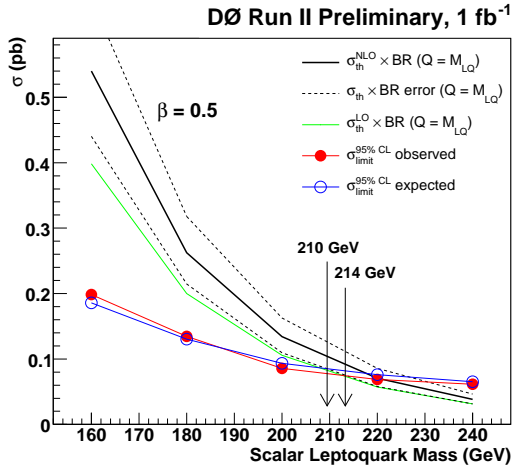}
\end{center}
\caption{Search for $LQ_2\overline{LQ}_2\ra \mu q \nu q$: comparison
of the upper cross-section limit for the pair-production of second
generation leptoquarks with the predicted cross section.}
\label{fig:lq2}
\end{figure}

Upper limits on the cross section times branching ratio were obtained
and compared to the NLO prediction reduced by its uncertainty to derive
a leptoquark mass limit as shown in Fig.~\ref{fig:lq2}.
From this analysis alone, a lower mass limit for scalar second generation
leptoquarks of $M_{LQ_2}>214\,\gev$ at $\beta=0.5$ was set. The
previous best limits were $M_{LQ_2}>170\,\gev$ obtained in the $\mu\nu jj$ 
channel and $M_{LQ_2}>208\,\gev$ from a combination with the $\mu\mu jj$ and
$\nu\nu jj$ channels~\cite{Abulencia:2005et2}.

\section{Single production of scalar leptoquarks}
The production of single leptoquarks leads to final states
consisting of two leptons and one jet.
A search for single leptoquark production was performed using $\mu\mu j$
events reconstructed in a data sample of $300\,\fbi$ \cite{Abazov:2006ej}.
Despite their couplings to the second lepton generation, the leptoquarks
were assumed to couple to the first quark generation, to avoid a
suppression of the cross section due to the parton distribution functions.
In the standard model there is no explicit connection between a given lepton 
generation with any of the three quark generations.
The dominant background in this search was found to be from $Z/\gamma^* +$ jets
production.

\begin{figure}
\begin{center}
\includegraphics[width=0.44\textwidth]{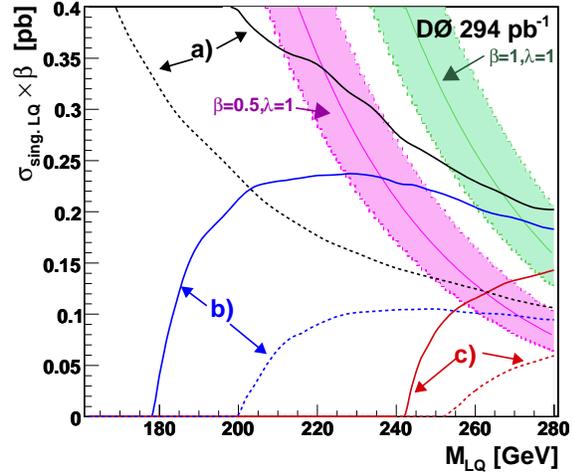}
\end{center}
\caption{Search for $\mu LQ\ra \mu\mu q$: cross section limits for the 
production of single leptoquarks for the three scenarios described in the 
text, compared to the theoretical prediction.}
\label{fig:1lq}
\end{figure}

The event distribution in the two-dimensional plane given by the di-muon mass
$M_{\mu\mu}$ and the $E_T$ of the leading jet was used to define four signal
bins. Those were combined with the three signal bins of the leptoquark
pair-production search in the $\mu\mu jj$ channel~\cite{Abazov:2006vc}, in 
order to derive upper cross section limits on the production of single
leptoquarks. The limits are given in Fig.~\ref{fig:1lq} for three different
scenarios: (a) the only contribution in the signal region is from SM background
and single leptoquark production, (b) pair-production contributes in addition
with the assumption $\beta= 1/2$, and (c) as (b) but with $\beta=1$.
Lower limits on the leptoquark mass were derived assuming $\lambda =1$. 
Compared to the search for leptoquarks which considered only pair-production
(corresponding to $\lambda \ll 1$) the mass limits were improved to
$M_{LQ}>274\,\gev$ for $\beta =1$ and to $M_{LQ}>226\,\gev$ for $\beta =1/2$,
respectively.

\section{Pair-production of third generation leptoquarks}
Searches for the pair-production of third generation leptoquarks were
performed in the $\tau b \tau b$ and $\nu b \nu b$ final states.

\subsection{The $\mathbold{\tau\tau bb}$ final state}
The search for $LQ_3\overline{LQ}_3\ra \tau b \tau b$ is based on the Run\,IIa
data set of $1\,\fbi$~\cite{d05447}. One of the taus was required to
decay into a muon ($\tau_\mu$) and the other tau needed to decay hadronically
($\tau_h$).
Hadronic $\tau$ decays were reconstructed from calorimeter clusters and tracks
and were separated into three types based on their decay. For each $\tau$ type
a neural network was used for the discrimination against background.
The $b$ quark jets were identified using a neural network tagging algorithm
with relatively loose conditions, corresponding to a $b$-tagging efficiency
of 72\% at a light quark misidentification rate of 6\%. Subsamples with one and
two $b$-tags, respectively, were defined.

\begin{figure}
\begin{center}
\includegraphics[width=0.45\textwidth,bb=0 0 560 370,clip]{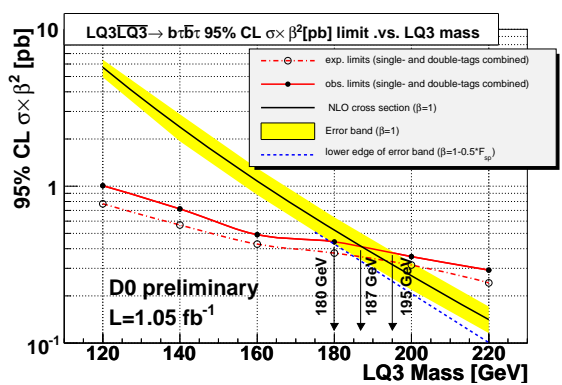}
\end{center}
\caption{Search for $LQ_3\overline{LQ}_3\ra \tau b \tau b$: observed and
expected upper limit on
the pair-production cross section compared to the theoretical prediction.}
\label{fig:lq3tau}
\end{figure}

The main background contribution were $t\bar{t}$, $Z/W +$ jets, and QCD
multijet production. The contribution from the latter was estimated using
like-sign $\tau_\mu\tau_h$ candidate events. The scalar sum of transverse energies
$S_T=E_T(\mu)+ E_T(\tau_h) + E_T(\mathrm{jet}_1) + E_T(\mathrm{jet}_2) + \MET$
was used as the main discriminant between the SM background and the leptoquark
signal. The dominant systematic uncertainties were due to
uncertainties in the cross sections of the background processes and in the
$b$-tagging efficiency.
Lower limits on the leptoquark mass were derived from the combination
of the single-tag and double-tag subsamples (see Fig.~\ref{fig:lq3tau}).
Assuming the hypothetical leptoquark has charge-$4/3$, which implies a 
branching fraction
$\mathcal{B}(LQ_3\ra \tau b)=1$, a lower mass limit $M_{LQ_3}>180\,\gev$
was set, which corresponds to an upper cross section limit of $0.42\,\pb$.
For charge-$2/3$ leptoquarks, decays into $\nu t$ are allowed as well, albeit
those are kinematically suppressed. Assuming equal leptoquark couplings to
$\tau b$ and $\nu t$, $\mathcal{B}(LQ_3\ra \tau b)$ hardly changes and the 
same mass limit was obtained.

Based on the same final state, the CDF collaboration previously published
a search for the pair-produc\-tion of
third generation {\em vector} leptoquarks~\cite{Aaltonen:2007rb}. When 
including the uncertainty on the signal cross-section prediction, they obtained
lower mass limits of $M_{LQ_3}>294\,\gev$ and $M_{LQ_3}>223\,\gev$ for the case
of Yang-Mills couplings and the minimal couplings model, respectively.
The latter corresponds to an upper cross-section limit of $0.61\,\pb$.

\subsection{The $\mathbold{bb{E \kern-0.6em\slash_{\rm T}}}$ final state}
The D0 collaboration recently published a search for 
the pair-production of third generation scalar leptoquarks
decaying into a neutrino and a $b$ quark using $425\,\pbi$ 
of data collected with a missing transverse energy and a
single-muon trigger~\cite{Abazov:2007bs}.
Two selections based on the two different triggers were combined in the 
analysis.
In both cases minimal requirements on the leading and
second-leading jet $E_T$ and on \MET{} were applied. Furthermore, two jets
were required to be tagged as $b$ jets with at least one of them by a
jet lifetime probability algorithm based on the impact-parameters of the 
associated tracks. For
the selection using the single-muon triggers, which is motivated by the
semi-leptonic decays of $B$-mesons, one of the jets was required to be tagged
with a soft-muon tagger.

\begin{figure}
\begin{center}
\includegraphics[width=0.45\textwidth]{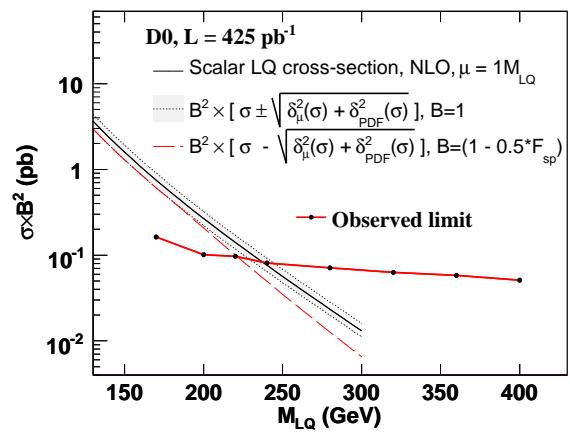}
\end{center}
\caption{Search for $LQ_3\overline{LQ}_3\ra \nu b \nu b$: upper limit on
the pair-production cross section compared to the theoretical prediction.}
\label{fig:lq3met}
\end{figure}

The cuts on the final selection variables, \MET{} and scalar 
$H_T=\sum_\mathrm{jets} E_T$, were optimized
as function of the assumed leptoquark mass $M_{LQ_3}$. 
The main background sources were determined to be
$t\bar{t}$, $W/Z+b\bar{b}$, and $W/Z+c\bar{c}$ production. 
The uncertainty
on their contribution was found to be dominated by uncertainties on the
cross-section predictions, on the jet energy calibration, and on the
$b$-tagging efficiency. Assuming that the leptoquarks have charge-{1/3} and 
that they decay exclusively in
a neutrino and a $b$ quark, a mass limit on third-generation leptoquarks
of $M_{LQ_3}>229\,\gev$ was derived as shown in Fig.~\ref{fig:lq3met}.
Taking into account leptoquark decays into $\tau$ and $t$ quark as well and
assuming equal couplings, a mass limit of $M_{LQ_3}>221\,\gev$ was set. 

\section{Conclusions and perspectives}
The D0 experiment at the Tevatron $p\bar{p}$-collider has searched for single and
pair-production of leptoquarks in a multitude of final states using data sets
with up to $1\,\fbi$. No indication
for leptoquark production has been found and stringent limits, 
which are significantly improved compared to Run\,I, were set.
Nearly $3\,\fbi$ of integrated luminosity has been recorded in summer 2007
and the experiment is expected to collect much larger data
sets during the full period of Run\,II.


\end{document}